\documentclass[letter]{aa} 

\usepackage{graphicx}
\usepackage{txfonts}
\definecolor{smalt(darkpowderblue)}{rgb}{0.0, 0.2, 0.6}
\definecolor{forestgreen(traditional)}{rgb}{0.0, 0.5, 0.0}

\defcitealias{RVJ}{RVJ}

\defcitealias{MD17}{MD17}

\newcommand{\cv}{cataclysmic variable}
\newcommand{\cvs}{cataclysmic variables}

\defcitealias{ACR}{ACR}

\begin{document}

   \title{The cataclysmic variable orbital period gap: More evident than ever} 


   \author{
          Matthias R. Schreiber\inst{1.2}
          \and
          Diogo Belloni
          \inst{1}
          \and 
          Axel D. Schwope \inst{3}
          }

    \authorrunning{Schreiber et al. }

  \institute{Departamento de F\'isica, Universidad T\'ecnica Federico Santa Mar\'ia, Av. España 1680, Valpara\'iso, Chile\\
              \email{matthias.schreiber@usm.cl}
         \and
             Millenium Nucleus for Planet Formation, Valpara{\'i}so, Chile
        \and
        Leibniz-Institut für Astrophysik Potsdam (AIP), An der Sternwarte 16, D-14482 Potsdam, Germany
        }

   \date{}

  \abstract
   {Recently, large and homogeneous samples of cataclysmic variables identified by the Sloan Digital Sky Survey (SDSS) were published. In these samples, the famous orbital period gap, which is a dearth of systems in the orbital period range $\sim2-3$\,hr and the defining feature of most evolutionary models for \cvs, has been claimed not to be clearly present. If true, this finding would completely change our picture of \cv~evolution. } 
   {In this Letter we focus on potential differences with respect to the orbital period gap between \cvs~in which the magnetic field of the white dwarf is strong enough to connect with that of the donor star, so-called polars, and non-polar \cvs~as the white dwarf magnetic field in polars has been predicted to reduce the strength of angular momentum loss through magnetic braking.  }
   {We separated the SDSS I-IV sample of \cvs~into polars and non-polar systems and performed statistical tests to evaluate whether the period distributions are bimodal as predicted by the standard model for \cv~evolution or not.  We also compared the SDSS\,I-IV period distribution of non-polars to that of other samples of \cvs.}
   {We confirm the existence of a period gap in the SDSS\,I-IV sample of non-polar \cvs~with $>98$ per cent confidence. The boundaries of the orbital period gap are $147$ and $191$ minutes, with the lower boundary being different to previously published values ($129$\,min). 
   The orbital period distribution of polars from SDSS I-IV is clearly different and does not show a similar period gap. }
   {The SDSS samples as well as previous samples of \cvs~are consistent with the standard theory of \cv~evolution. Magnetic braking does indeed seem get disrupted around the fully convective boundary, which causes
   a detached phase during \cv~evolution. In polars, the white dwarf magnetic field reduces the strength of magnetic braking and consequently the orbital period distribution of polars does not display an equally profound and extended period gap as non-polars. It remains unclear why the breaking rates derived from the rotation of single stars in open clusters favour prescriptions that are unable to explain the orbital period distribution of \cvs.} 

   \keywords{
   binaries: close --
             stars: evolution --
             stars: novae, cataclysmic variables --
             white dwarfs
            }
   \maketitle
%

\section{Introduction}

According to the standard scenario for the evolution of cataclysmic variables, angular momentum loss due to magnetic braking is orders of magnitude stronger than gravitational radiation if the secondary stars still contain a radiative core. The correspondingly large mass transfer rates drive the donor stars out of thermal equilibrium. As soon as the donor star becomes fully convective, at about an orbital period of $\sim3$\,hr, magnetic braking becomes much less efficient, the donor star has time to relax, and the system becomes a detached binary until reduced magnetic braking and gravitational radiation bring the stars close enough to restart mass transfer (at a much lower rate) at an orbital period of $\sim2$\,hr \citep[e.g.][]{kolb93-1,belloni+schreiber23-1}. As this evolutionary scenario includes a drastic decrease in
magnetic braking at the fully convective boundary, it is known as disrupted magnetic braking. 

The main motivation for developing the previously described scenario has been 
the observed dearth of \cvs~in the period range between $\sim2-3$\,hr,  
the so-called period gap. However, whether observed samples do indeed provide statistically significant evidence for a reduction in \cvs~with orbital periods between $\sim2-3$\,hr and whether this perhaps depends on the considered sub-type of \cvs\  has been intensively discussed during the last decades. 

\citet{verbunt97-1} claimed that the gap is not statistically significant for \cvs~with the highest mass transfer rates (so-called nova likes), which was refuted by \citet{hellier+naylor98-1} and \citet{kolbetal98-1}. 
\citet{hellier+naylor98-1}, as well as \citet{webbink+wickramasinghe02-1}, argued that polars, that is, \cvs~in which the white dwarf magnetic field connects with that of the donor star that synchronises the white dwarf spin and the orbit, should be less affected by magnetic braking and therefore should show a shorter and less pronounced period gap. 
While \citet{wheatley95-1} found no evidence for a difference in the period distribution of polars and non-polars, more recent works show that 
the gap in the polar distribution of the Ritter \& Kolb catalogue \citep{ritter+kolb03-1} of \cvs~is less pronounced than that 
for non-magnetic \cvs~\citep{pretoriusetal13-1}. 

A recently established volume-limited sample of \cvs~seems to show a period gap but is subject to low-number statistics \citep{palaetal20-1}. In contrast, the period distribution of the large and homogeneous sample of \cvs~spectroscopically discovered by the Sloan Digital Sky Survey (SDSS) 
has been claimed to not show clear evidence for a period gap \citep{inightetal23-2}.   

As magnetic braking is not only of crucial importance for \cvs~but also for other binary stars, and as it also drives the spin down of single low-mass stars, one might have hoped that independent observational or theoretical constraints 
could settle the discussion. Unfortunately, this has not been the case. 

On the one hand, from \cvs~and related objects, there seems to be strong 
support for disrupted magnetic braking from other observables than the period distribution of \cvs~\citep{schreiberetal10-1,kniggeetal11-1,zorotovicetal16-1,mcallisteretal19-1}. 
On the other hand, observations of single stars as well as main sequence binaries argue against the disrupted magnetic braking scenario
\citep[e.g.][]{gossageetal23-1, elbadryetal22-1}, and favour saturated magnetic braking prescriptions instead. 
According to these saturated magnetic braking prescriptions, the relation between angular momentum loss and the spin period of the mass losing star saturates for rotation periods below a critical value in a similar fashion as chromospheric activity, coronal X-ray emission, and flare activity saturate \citep[e.g.][]{reinersetal09-1,newtonetal17-1,johnstoneetal21-1}. 

Unfortunately, most saturated magnetic braking prescriptions proposed so far predict angular momentum loss rates that are too weak to explain the orbital period gap, that is, the resulting mass transfer rates are not high enough to 
sufficiently increase the radius of the donor star. 
This means that, if the period gap exists, we need much stronger magnetic braking in \cvs~than what is usually assumed for single main sequence stars or main sequence star binaries. 
If, instead, the period gap does not exist, most evidence would point towards a universal saturated magnetic braking prescription and we would need different explanations for the above listed independent evidence for disrupted magnetic braking. 
It is thus important to determine whether the period gap is a real feature (and, if so, for which type of \cv) or if it just appeared in early samples of \cvs~due to observational biases and selection effects as suggested by \citet{inightetal23-2}.

\section{The SDSS I-IV sample} 

\begin{figure}
        \centering
    \includegraphics[width=1\columnwidth]{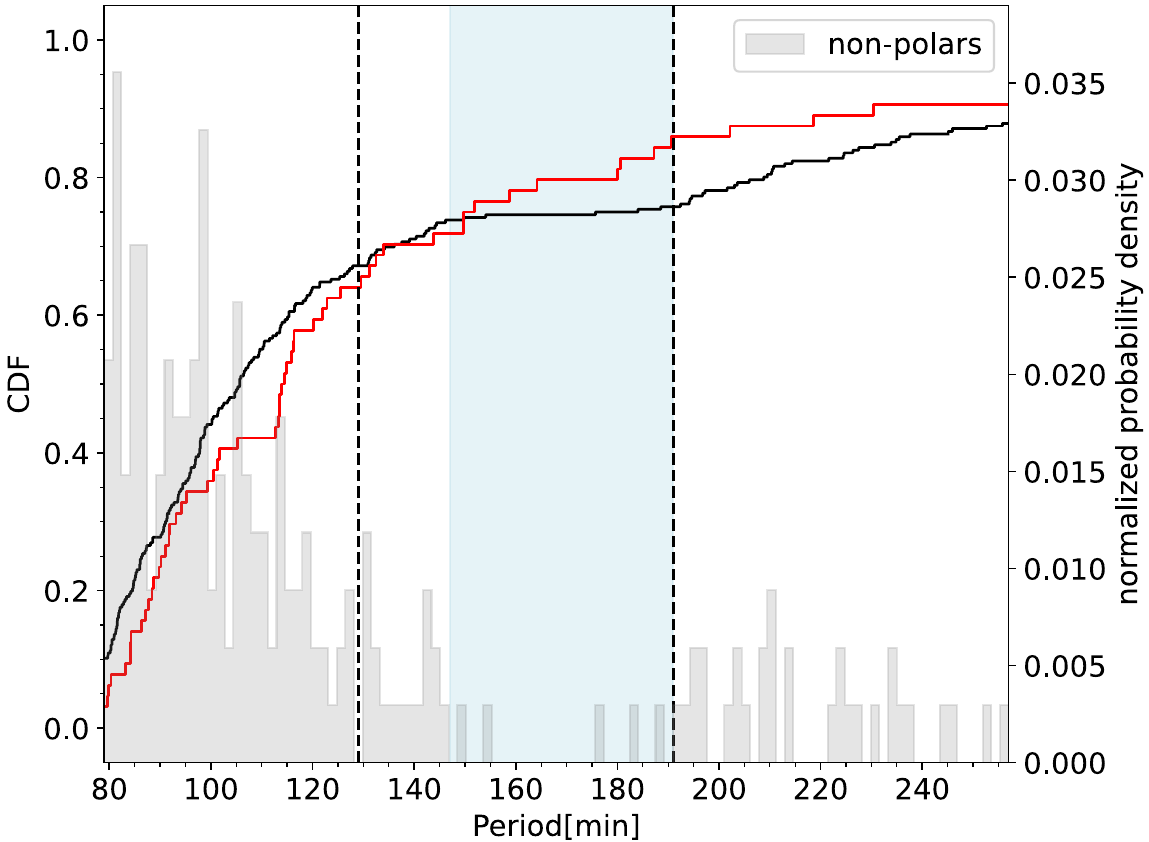}
    \caption{
    Period distribution of both polars (red line) and non-polar \cvs~(black line and grey histogram) from the SDSS I-IV sample of \cvs~recently published by \citet{inightetal23-1}. A dearth of systems between 147 and 191 minutes (light blue shaded region) in the sample of non-polars can clearly be detected. We claim that the reduced number of non-polar \cvs~in this period range represents the (in)famous period gap of \cvs. The dashed vertical lines indicate the boundaries of the period gap as identified by \citet{kniggeetal11-1}, while the upper boundary perfectly fits with what we found here, the lower edge of the period gap in the SDSS I-IV sample seems to be located at longer periods. The period distribution of polars does not show evidence for a reduction in the number of systems in the gap.}
    \label{fig:cum}
\end{figure}

\begin{figure}
\centering
\includegraphics[width=0.95\linewidth]{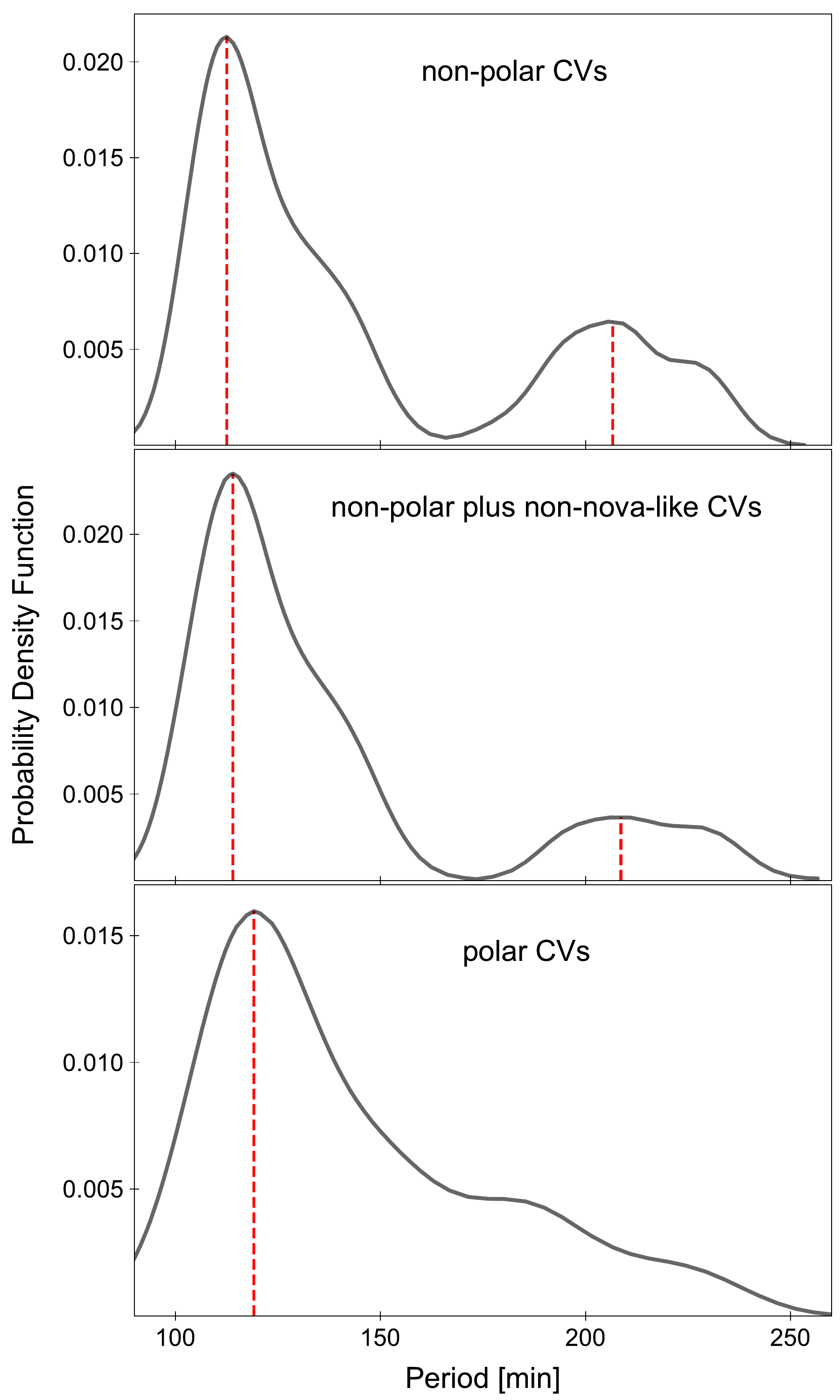}
\caption{Probability density functions based on kernel density estimations with the Gaussian kernel and adopting the critical bandwidth of \citet{Silverman1981} of the period distribution of different sub-samples of \cvs. From the top to bottom panels, we excluded the polars, the polars plus the nova likes, and finally the non-polars. The vertical dashed lines indicate the location of the modes assuming bimodal distributions (top and middle panels) and a unimodal distribution (bottom panels).}
\label{fig:kde}
\end{figure}

In two extensive works, \citet{inightetal23-1,inightetal23-2} recently established large samples of \cvs~identified by SDSS and convincingly showed that the orbital period distributions of these more homogeneous samples are significantly different to that of the Ritter \& Kolb catalogue. 
In this work, we mainly consider the SDSS\,I-IV sample as \cvs~in this catalogue are all serendipitous identifications through low-resolution SDSS spectroscopy.
This sample represents the largest homogeneous sample of \cvs~currently available. 
SDSS spectroscopy has been crucial for \cv~studies as, for example, the large number of low-accretion rate systems below the gap and at the orbital period minimum, which had been predicted for a long time, was only discovered thanks to early SDSS samples of \cvs~\citep{gaensickeetal09-1}. 
 
Following the idea of \citet{webbink+wickramasinghe02-1}, \citet{bellonietal20-1} showed that the period distributions predicted for polars and for the rest of the \cv~population (from now on non-polar \cvs) should be different with respect to the gap because in polars the white dwarf magnetic field reduces the wind zones of the donor star and thereby the efficiency of magnetic braking. According to this prediction, the mass transfer rates above the gap should be lower, 
and the donors should be less (if at all) driven out of thermal equilibrium. Without a significantly inflated donor star at long orbital periods (above the gap), 
a much shorter or no detached phase is expected to be generated for polars by disrupted magnetic braking. 

We therefore show in Fig.\,\ref{fig:cum} the cumulative distributions of polars and non-polar \cvs~separately. 
It becomes immediately clear that in the period range where the period gap has been claimed to exist, the period distribution of polars differs from that of non-polars.
Only the latter ones show clear evidence for a period gap.
Looking at the cumulative distributions and using different binning for histograms, we identified periods between $147$ and $191$ minutes by eye
as the region with the most obvious decrease in the number of non-polar \cvs~per period interval. 
Only five of the 256 non-polar \cvs, which corresponds to $2.0$\,per cent, have periods in this range. 
The lower boundary of the gap ($147$ min.) is different to the value ($129$ min.) previously found by \citet{kniggeetal11-1}. 
The period distribution of polars does not show an obvious gap, that is, a reduction in the number of systems in a certain period range. In the range of periods we defined as the gap for non-polar \cvs~($147-191$ min), we found nine out of 64 polars ($14.1$ per cent). 
In what follows, we explain how we derived
the statistical significance of the previously described observations. 

\begin{table}
\caption{Characteristics of the probability density derived using the kernel density estimation method adopting Gaussian kernels with the critical bandwidth calculated according to \citet{Silverman1981}.}
\label{table:1}
\setlength\tabcolsep{6pt} 
\renewcommand{\arraystretch}{1.2} 
\centering
\begin{tabular}{lrrr}
\hline\hline
Property & polars & \multicolumn{2}{c}{non-polars} \\
 & & with NL &  without NL \\
\hline
critical bandwidth (min)                        &  12.91  &    6.59  &    7.48  \\
\multicolumn{4}{l}{mode 1} \\
location (min)                  & 119.36  &  111.54  &  112.05  \\
density ($10^{-2}$ min$^{-1}$)  &   1.60  &    2.12  &    2.34  \\
\multicolumn{4}{l}{mode 2} \\
location (min)                  &         &  206.40  &  208.14  \\
density ($10^{-3}$ min$^{-1}$)  &         &    6.49  &    3.60  \\
\multicolumn{4}{l}{antimode} \\
location (min)                  &         &  165.90  &  172.93  \\
density ($10^{-5}$ min$^{-1}$)  &         &   39.29  &    6.62  \\
\hline
\end{tabular}
\tablefoot{NL: nova like}
\end{table}

\section{The statistical significance of the period gap}
\label{sec:sig}

To determine the statistical significance of the decrease in the number of non-polar \cvs~in the period range between 147 and 191 minutes, we performed several unimodality and bimodality tests using only systems with periods between 103 and 235 minutes (as this range is sufficient to cover the period gap).
If the observed decrease in the number of \cvs~is statistically significant, the period distribution around the gap should be consistent with a bimodal distribution.

As in the previous section, we separated polars and non-polars. Given that even for the homogeneous SDSS\,I-IV sample of \cvs,~observational biases cannot be excluded, we defined an additional sample excluding nova likes from the non-polars. Nova likes are \cvs~with stable accretion disks, have the largest mass transfer rates, and are typically found just above the period gap. If observational biases were dominating, these systems would be overrepresented which could in principal explain the period gap in observed magnitude-limited samples. By eliminating nova likes from the non-polar sample, we therefore tested how strong observational biases towards the detection of nova likes could impact our results on the statistical significance of the period gap. All samples were established based on the classifications by \citet{inightetal23-1}.

For each of the three samples, we first applied unimodality tests and, in the case the null hypothesis (the true number of modes is one) can be rejected with more than $95$ per cent confidence, we performed bimodality tests.
We used the tests suggested by \citet{Silverman1981} and \citet{ACR}
as provided in the multimode package implemented by \citet{Ameijeiras_Alonso_2021} in the numerical tool R \citep{R}.
For the \citet{Silverman1981} test, we first determined the critical bandwidth and then generated resamples using the distribution associated with the corresponding kernel density estimation. 
In the case of the \citet{ACR} test, 
we used the exact excess mass value to perturb the sample data. For each test, we chose three different values for the number of bootstrap replicates (100, 500, and 1000).

The results of our tests are as follows.
Regarding the sub-sample of non-polars,
we can reject the null hypothesis (the true number of modes is equal to one) with at least 98.0
and at least 99.6 per cent
confidence for the \citet{Silverman1981} and the \citet{ACR} test, respectively.
This strong evidence for the existence of the period gap slightly decreases if the brightest \cvs~(nova likes) are excluded, but it remains above 95 and 96.6 per cent. 
Performing bimodality tests for both samples, we find that the null hypothesis 
(i.e. that the distribution is bimodal)
cannot be rejected ($p$ values exceeding 20 per cent). 
Consequently, the period distribution 
of non-polars is bimodal and this statement remains true when nova likes are excluded. 

In contrast, unimodal tests for polars provide $p$ values largely exceeding five per cent, which means the null hypothesis (the true number of modes is equal to one) cannot be rejected. 
However, 
given the small sample size of polars (only 39 polars have periods between $103$ and $235$ minutes), 
we cannot exclude the existence of 
a less pronounced gap similar to the one found in the Ritter \& Kolb catalogue 
\citep{pretoriusetal13-1,schwopeetal20-1}.
From these test results, we conclude that 
(i) there is a statistically significant gap in the period distribution of non-polar \cvs~from SDSS\,I-IV; (ii) this gap is not caused by observational biases favouring the detection of 
nova-like \cvs; and (iii)  
there is no statistically significant evidence for the existence of a similar period gap in the distribution of polars from SDSS\,I-IV. 

We show in Fig.~\ref{fig:kde} the probability density functions derived with the kernel density estimation method adopting Gaussian kernels and the critical bandwidth determined according to \citet{Silverman1981}. 
In line with the test results described above, we show a bimodal distribution for the non-polar samples and a unimodal distribution for the polar sample.
The locations of the modes and the corresponding densities are given in Table \ref{table:1}. 

\section{Comparison with other samples}
\label{app:com}

The SDSS\,I-IV sample presented by \citet{inightetal23-1} represents the largest sample of \cvs~identified in a homogeneous way (SDSS spectroscopy). 
In what follows we compare the period distribution from SDSS I-IV with those of other recently established catalogues of \cvs. 
For this exercise we only consider non-polars because 
the number of polars in the other samples is too small to draw any meaningful conclusions concerning the polar period distribution. 

\begin{figure}
        \centering
    \includegraphics[width=1\columnwidth]{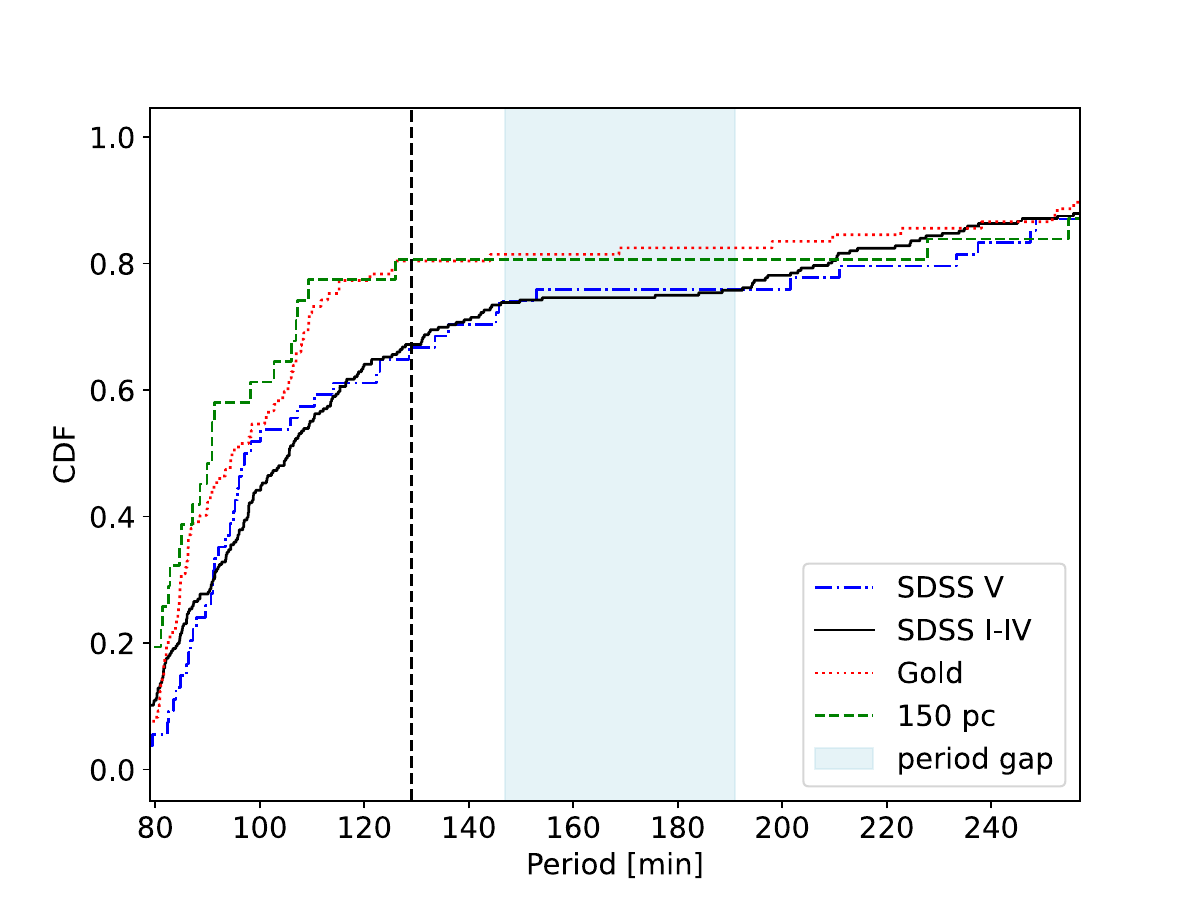}
    \caption{Period distributions of non-polar \cvs~from SDSS I-IV \citep{inightetal23-1}, the plate survey that is part of SDSS V \citep{inightetal23-2}, the incomplete but volume-limited (300\,pc) Gold sample \citep{inightetal21-1}, as well as the largely complete but small 150\,pc sample \citep{palaetal20-1}. 
    While both SDSS samples show a clear period gap in the range between $147$ and $191$ min, the volume-limited samples contain too few \cvs~with periods exceeding $\sim120$ min to derive meaningful 
    constraints concerning the period gap. This result illustrates that the SDSS samples are still significantly biased against \cvs~with periods shorter than two hours. 
    }
    \label{fig:samples}
\end{figure}

Figure \ref{fig:samples} shows the cumulative distributions of non-polar \cvs~from 
SDSS I-IV \citep{inightetal23-1}, the plate survey that is part of SDSS V \citep{inightetal23-2}, the incomplete but volume-limited (300\,pc) Gold sample \citep{inightetal21-1}, as well as the largely complete volume-limited 150\,pc sample \citep{palaetal20-1}.
The period distributions from the two SDSS samples are very similar (the KS test provides a $p$ value of $0.50$).   
Also the two volume-limited samples seem to agree with each other, that is, show no evidence to be not drawn form the same parent sample ($p$ value of $0.75$). 
However, we clearly observe a difference between the SDSS and the volume-limited samples ($p$ values below $0.09$ and below $0.02$ for comparison with SDSS\,V and SDSS\,I-IV, respectively). 

The differences arise from the fact that the volume-limited samples are more dominated by short orbital period \cvs~and contain few non-polar \cvs~with periods longer than $120$ min. 
Because of this, 
both volume-limited samples do not provide strong evidence in favour or against the existence of the period gap nor do they allow
one to distinguish between the lower gap boundary as defined by \citet{kniggeetal11-1} using the Ritter \& Kolb catalogue (129 min) and the one we found in the SDSS samples (147 min). 
However, the relative number of non-polar systems in the period gap (defined as 147-191 min) is similarly small in all samples, that is,
$0/31$, $1/97$, $1/55$, and $5/256$ for the 150\,pc, Gold, SDSS\,V, and SDSS\,I-IV samples, respectively.

The comparison between the different samples shows 
that all currently known samples containing large numbers of non-polar \cvs~with periods exceeding 120 min (i.e. SDSS I-IV, SDSS V, Ritter \& Kolb) show a period gap. Current volume-limited samples do not contradict this finding but provide no additional constraints.
The difference between the period distributions of volume-limited samples and that of the SDSS samples indicate the presence of observational biases affecting the SDSS samples. 

\section{Potential biases}

First of all, we do not see any reason why non-polar \cvs~should be more difficult to detect when they have periods between 147 and 191 minutes than non-polar \cvs~below the gap. The only observational bias one could imagine would be a drastic decrease in the mass transfer rate, the extreme case of which is exactly what disrupted magnetic braking predicts (no mass transfer at all). 
As shown in Section\,\ref{sec:sig}, the gap remains to be statistically significant even if the brightest \cvs~(nova likes) are excluded from the non-polar sample. In addition, the fraction of \cvs~is similarly low in the SDSS and the volume-limited surveys. All this shows that the gap is extremely unlikely to be caused by observational biases.

We also do not see any reason why polars should be easier to detect in the period gap than non-polar \cvs, which implies that the difference between the two subgroups in the SDSS I-IV sample is real. This shows that the existence of the gap is related to magnetism which strongly supports the idea of disrupted magnetic braking.  

However, compared to the volume-limited samples, the SDSS samples contain significantly fewer systems below the period gap, that is, the SDSS samples are still biased against short period low mass transfer systems. This bias might affect the location of the lower boundary of the gap. 
A large volume-limited sample of \cvs, which perhaps can be provided by combining eROSITA detections with suitable follow-up observations, is needed to finally 
determine the exact location and depth of the gap from a fully representative sample. 

However, it is fundamentally important to note that, for the time being, the exact location of the period gap does not provide critical information. Individual evolutionary tracks of \cvs~predict the detached phase and the restart of mass transfer at slightly different orbital periods depending on the initial donor mass and metallicity as well as the white dwarf mass. 
Some \cvs~start mass transfer in the gap and are not affected by disrupted magnetic braking. Furthermore, the appearance of white dwarf magnetic fields during \cv~evolution can alter the evolution of individual systems 
\citep{schreiberetal21-1}. A razor-sharp orbital period gap is thus not expected and one should not search for it. 

Once a large and volume-limited sample of \cvs~has been established, one might be able to extract further information on magnetic braking by comparing the exact shape of the period distribution with theoretical predictions. In the absence of this large volume-limited sample, the key information we can retrieve from the currently available data is the following: 
all existing samples of \cvs~are consistent with a deep period gap in the distribution 
of non-polar \cvs~and a less pronounced (or no) gap in the period distribution of polars.

\section{Conclusion}

The Ritter \& Kolb catalogue has provided evidence for a period gap in the \cv~orbital period distribution. The existence of this period gap formed the basis for developing the theory of disrupted magnetic braking. 
In contrast to recent claims, the orbital period distributions of \cvs~from SDSS samples provide further evidence for disrupted magnetic braking. A clear period gap is present in the distribution of non-polar \cvs~and a similarly profound and extended gap can be excluded for polars from SDSS. A consistent picture for \cv~evolution 
therefore remains to contain the following ingredients: 

(i) Magnetic braking above the gap is stronger than predicted by the models of saturated magnetic braking. 

(ii) Magnetic braking needs to decrease significantly around the fully convective boundary.

(iii) Magnetic braking is reduced in polars as the wind from the secondary star is trapped within the magnetosphere of the white dwarf \citep[e.g.][]{bellonietal20-1}.

Despite these crucial results on magnetic braking (or actually rather because of them), it might be difficult to soon find a unified magnetic braking prescription. 
Such a universal magnetic braking prescription would need to convincingly explain observations of single stars, which seem to favour some kind of saturated magnetic braking \citep{gossageetal23-1}, reproduce the donor star radii in \cvs~as well as
disrupted magnetic braking 
\citep{kniggeetal11-1}, and provide the strong angular momentum loss rates 
that are required to explain persistent low-mass X-ray binaries \citep{van+ivanova19-1} and AM\,CVn binaries descending from \cvs~with evolved donors \citep{belloni+schreiber23-2}. 

A good starting point might be a disrupted and saturated magnetic braking prescription, which could work for \cvs~as well as detached main sequence and post common envelope binaries \citep{bellonietal23-1}. However, this alone would not solve the differences between the strength of magnetic braking in \cvs, the progenitors of low-mass X-ray binaries, and single stars. Potentially new dependencies on the age and evolutionary status of the stars need to be considered. 

It appears unlikely that the large mass transfer rates derived for \cvs~above the gap from the radii of their donors \citep{mcallisteretal19-1}, and partly also from the white dwarf temperatures \citep{palaetal22-1}, 
are caused by consequential angular momentum loss instead of magnetic braking as speculated by \citet{elbadryetal22-1}. Consequential angular momentum loss might play a role in explaining the white dwarf mass distribution of \cvs~\citep{schreiberetal16-1}, but it can hardly be the main angular momentum loss mechanism above the gap. This is because there is no obvious reason for consequential angular momentum loss to be less efficient in polars and/or to decrease at the fully convective boundary. 

As a final remark, while all samples of the \cvs~currently available agree with the main predictions of disrupted magnetic braking, important issues related to our understanding of \cv~evolution still need to be addressed.
On the observational side, a large and complete volume-limited sample is not yet available and we therefore cannot fully exclude that observational biases play a role. Concerning theoretical predictions, recent evidence points towards the appearance of white dwarf magnetic fields during \cv~evolution, which can lead to significant deviations from the standard scenario of \cv~evolution 
\citep{schreiberetal21-1}. A binary population
synthesis code including the predicted transition from non-polar \cvs~to polars is not yet available.  
Last but not least, the fact that large numbers of \cvs~that passed the period minimum are predicted to exist but are absent in observed samples remains puzzling \citep{inightetal23-2}. 
Perhaps the latter two issues are related as the late appearance of white dwarf magnetic fields may generate extended detached phases in period bouncers \citep{schreiberetal23-1}.

\begin{acknowledgements}
We thank B.T. G\"ansicke and K. Inight for lively discussions and for providing the SDSS samples of \cvs. 
This research was supported by the National Science Foundation under grant No. NSF PHY-1748958 
and the Deutsche Forschunsgemeinschaft (DFG) under grants EXC-2094–390783311 and Schw536/37-1.
MRS and DB are supported by FONDECYT (grant numbers 1221059 and 3220167). MRS further acknowledges support from ANID, – Millennium Science Initiative Program – NCN19\_171.
%
%

\end{acknowledgements}

%
%

\bibliographystyle{aa} 
\bibliography{gap_evidence} 

\end{document}